\documentstyle[12pt,epsfig,amstex]{article}
\chardef\anciennecat=\catcode`\@
\catcode`\@=11
\renewcommand\section{\@startsection {section}{1}{\z@}%
                                   {-3.5ex \@plus -1ex \@minus -.2ex}%
                                   {2.3ex \@plus.2ex}%
                                   {\reset@font\large\bfseries}}
\catcode`\@=\anciennecat

\DeclareSymbolFont{AMSb}{U}{msb}{m}{n}
\DeclareMathSymbol{\C}{\mathalpha}{AMSb}{"43}
\DeclareMathSymbol{\R}{\mathalpha}{AMSb}{"52}
\DeclareMathSymbol{\Pom}{\mathalpha}{AMSb}{"50}
\DeclareMathSymbol{\T}{\mathalpha}{AMSb}{"54}

\begin{document}

\title{\bf Pomeron intercepts at colliders }
\author{R. Peschanski {\it (SPhT)}\\
Ch. Royon {\it (DAPNIA/SPP)}\\
--- \\
 {\it Commissariat \`a l'Energie Atomique, Saclay}, \\
 {\it F-91191 Gif-sur-Yvette Cedex}\\
{\it France}
}
\maketitle

\begin{abstract}
A method allowing for a direct comparison of data with theoretical  predictions
is 
proposed for forward jet 
production at HERA and Mueller-Navelet jets at Tevatron and LHC.
An application to the determination of the {\it effective } Pomeron intercept in
the 
BFKL-LO parametrization from $d\sigma/dx$ data at HERA   leads to a good fit
with a 
significantly higher {\it effective }
intercept, $\alpha_P= 1.43 \pm 0.025 (stat.) \pm 0.025 (syst.),$ than for proton
(total 
and diffractive) structure functions. It is however less than the value of the
pomeron
intercept using dijets with large rapidity intervals obtained at Tevatron.
We also evaluate the rapidity veto 
contribution to the higher order BFKL corrections. We suggest to measure 
the dependence of the dijet cross-sections as a function of the jet
transverse energies as a signal for BFKL pomeron at LHC.
\end{abstract}

\maketitle

\section{Forward jet cross-section at HERA}
The study of forward jets at colliders is considered as the milestone of QCD
studies 
at high energies,
since it provides a direct way of testing the perturbative resummations of soft
gluon 
radiation. More 
precisely, the  study of one forward jet (w.r.t. the proton) in an
electron-proton 
collider~\cite{mu91}
seems to be a good candidate to test the energy dependence of hard QCD 
cross-sections. It is similar to the previous proposal of studying two jets
separated by 
a large rapidity interval in hadronic 
colliders~\cite{mu86}, for which only preliminary results are
available~\cite{D0}. This 
test is also possible in $\gamma^*$-$\gamma^*$ scattering~\cite{ro99} but here
the 
statistics and the energy range are still insufficient to get a reliable
determination 
of the physical parameters for hard QCD 
cross-sections.
Indeed, the 
proposed (and favored for the moment being) set-up~\cite{mu91} is to consider
jets with 
transverse momentum $k_T$ of the  
order of the 
photon virtuality $Q$ allowing to damp the  QCD evolution as a function of $k_T$
(DGLAP 
evolution~\cite{al77}) in favor of the   evolution in energy at fixed $k_T$
(BFKL 
evolution~\cite{li77}).

In contrast to full Monte-Carlo studies we want to focus on the jet cross section 
$d\sigma/dx$ observable itself, 
by a consistent treatment of 
the 
experimental cuts and minimizing the uncertainties for that particular 
observable. Let us remark that 
our 
approach is not intended to provide a substitution to the other methods, since
the 
Monte-Carlo simulations have the 
great 
merit of 
making a  set of predictions for various observables. Hence, our method has to
be 
considered as complementary
to the others and dedicated to a better  determination of the {\it effective}
Pomeron 
intercept using the  $d\sigma/dx$ data. As we shall see, it will fix more
precisely this
parameter, but it will leave less constrained other interesting parameters, such
as the 
cross-section normalization.

The cross-section for forward jet production at HERA in the dipole model 
reads~\cite{ba92}:
\begin{eqnarray}
\frac{d^{(4)} \sigma}{dx dQ^2 dx_J d k_T^2 d \Phi} &=&
\frac{ \pi N_C \alpha^2 \alpha_S(k_T^2)}{Q^4 k_T^2} 
\ f_{eff} (x,\mu_f^2)
\ \Sigma e_Q^2 
\int_{\frac 12- i\infty}^{\frac 12+ i\infty} \frac{d \gamma}{2i \pi} 
\left( \frac{Q^2}{k_T^2} \right)^{\gamma} \times\ \ \ \ \ \ \ \ \ \ \ \ \ \ \ \
\ \ 
\nonumber \\
&\times &\ 
\exp \{\epsilon (\gamma,0) Y\} \left[ \frac{h_T(\gamma) +h_L(\gamma)}{\gamma}
(1-y) + \frac{h_T(\gamma)}{\gamma} 
\frac{y^2}{2} \right]
\label{dsigma}
\end{eqnarray}
where
\begin{eqnarray}
Y &=& \ln \frac{x_J}{x} \\
\epsilon(\gamma,p)&=& \bar{\alpha} \left[ 2 \psi(1) -\psi(p+1-\gamma)
-\psi (p+\gamma) \right] \\
f_{eff} (x, \mu_f^2) &=& G(x,\mu_f^2) + \frac{4}{9} \Sigma (Q_f+ \bar{Q_f}) \\
\mu_f^2 &\sim& k_T^2\ ,
\end{eqnarray}
are, respectively, $Y$ the rapidity interval between the photon probe and the
jet,
$\epsilon(\gamma,p)$ the BFKL kernel eigenvalues, $f_{eff}$ the effective
structure 
function 
combination, and $\mu_f$ the corresponding factorization scale.  
The main BFKL parameter is $\bar{\alpha},$ which is the (fixed) value of the
effective 
strong coupling constant in LO-BFKL formulae. Note that we gave
the 
BFKL formula not including the azimuthal dependence as we will stick to the  
azimuth-independent contribution with the dominant $\exp \{\epsilon (\gamma,0)
Y\}$ 
factor.

The so-called ``impact factors''
\begin{eqnarray}
\label{defh}
\left(\begin{array}{c}
h_T \\ h_L \end{array} \right) = \frac{\alpha_S (k_T^2)}{ 3 \pi \gamma} 
\frac{(\Gamma(1 - \gamma) \Gamma(1 + \gamma))^3}{\Gamma(2 - 2\gamma) \Gamma(2 +
 2\gamma)} \frac{1}{1 - \frac{2}{3} \gamma} \left( \begin{array}{c} (1 +
 \gamma)
(1 - \frac{\gamma}{2}) \\ \gamma(1 - \gamma) \end{array} \right)\ ,
\end{eqnarray}
are obtained from the $k_T$ factorization properties~\cite{ca91} of the coupling
of 
the BFKL amplitudes to external hard probes. The same factors can be related 
to the photon wave functions~\cite{bj71,mu98} within the equivalent context of
the QCD 
dipole model~\cite{mu94}.

Our goal is to compare as directly as possible the theoretical parametrization 
(\ref{dsigma}) to the 
data 
which are collected in experiments~\cite{h199,ze99}. The crucial point is how to
take 
into account the 
experimentally defined kinematic cuts \cite{h199,ze99}. 

The main problem to solve is to investigate the effect of these cuts on the 
determination of the 
integration variables leading to a prediction for $d\sigma/dx$ from the given 
theoretical formula for 
$d^{(4)} \sigma$ as given in formula (\ref{dsigma}). The effect is expected to
appear as 
bin-per-bin {\it 
correction factors} to be multiplied to the theoretical cross-sections for
average 
values of the 
kinematic variables for a given $x$-bin before comparing to data (e.g. fitting
the 
cross-sections) \cite{ourpap}.

 The experimental correction factors have been determined using
a toy Monte-Carlo designed as follows. We generate flat distributions in the
variables 
$k_T^2/
Q^2$, $1/ Q^2$,
$x_J,$ using reference intervals   which include the whole of the experimental 
phase-space (the $\Phi$ 
variable is not used in the generation since all the cross-section measurements
are $\phi$ independent). In practice, we get the correction factors by counting
the 
numbers of events 
which fulfill
the experimental cuts given in Table {\bf I} for each $x$-bin. The correction
factor is
obtained by the ratio to the number of events which pass the experimental cuts 
and the 
kinematic
constraints, and the number of events which fullfil only the kinematic 
constraints,i.e.
the so-called reference bin. The correction factors are given in reference
\cite{ourpap}.

Weperform a 
fit to the H1 and ZEUS data with only two free parameters. these are the {\it
effective} 
strong 
coupling
constant in LO BFKL formulae $\bar{\alpha}$ corresponding to the {\it effective}
Lipatov 
intercept
$\alpha_P= 1+4 \log 2 \bar{\alpha} N_C/\pi$, and the cross-section
normalisation. The 
obtained values
of the parameters and the $\chi^2$ of the fit are given in Table {\bf III} for a
fit to 
the H1 and ZEUS 
data
separately, and then to the H1 + ZEUS data together.

\begin{center}
\begin{tabular}{|c|c|c|c|c|c|c|} \hline
 fit & $\bar{\alpha}$ & $\alpha_P$ & Norm. & $\chi^2 (/dof)$ \\ 
\hline\hline
 H1 & 0.17 $\pm$ 0.02 $\pm$ 0.01 & 1.44 $\pm$ 0.05 $\pm$ 0.025 & 29.4 $\pm$ 4.8
 $\pm$ 
5.2 & 5.7 (/9)\\
 ZEUS & 0.20 $\pm$ 0.02 $\pm$ 0.01 & 1.52 $\pm$ 0.05 $\pm$ 0.025 & 26.4 $\pm$
 3.9 $\pm$ 
4.7 & 2.0 (/2)\\
 H1+ZEUS & 0.16 $\pm$ 0.01 $\pm$ 0.01 & 1.43 $\pm$ 0.025 $\pm$ 0.025 & 30.7
 $\pm$ 2.9 
$\pm$ 3.5 & 12.0 (/13)\\
D0 & 0.24 $\pm$ 0.02 $\pm$ 0.02 & 1.65 $\pm$ 0.05 $\pm$ 0.05 & & \\ 
\hline
\end{tabular}
\end{center}
\vskip .5cm
\begin{center}
{Table I- Fit results}
\end{center}

The $\chi^2$ of the fits have been calculated using statistical error only and
are very 
satisfactory
(about $0.6 \ per \ point$ for H1 data, and $1. \ per \ point$ for ZEUS data).
We give both statistical and systematic errors for the fit parameters. 
The values of the Lipatov intercept are close to one another and compatible
within errors for the H1 and ZEUS sets of data, and indicate a preferable medium
value 
($\alpha_P=1.4-1.5$). We also notice that the ZEUS data have
the tendency to favour a higher exponent, but the number of data points
used in the fit is much smaller than for H1, and the H1 data are also at
lower $x$. The normalisation
is also compatible between ZEUS and H1. The fit results are shown in Figure 1
and 
compared with
the H1 and ZEUS measurements.

\begin{figure}
\begin{center}
\centerline{\psfig{figure=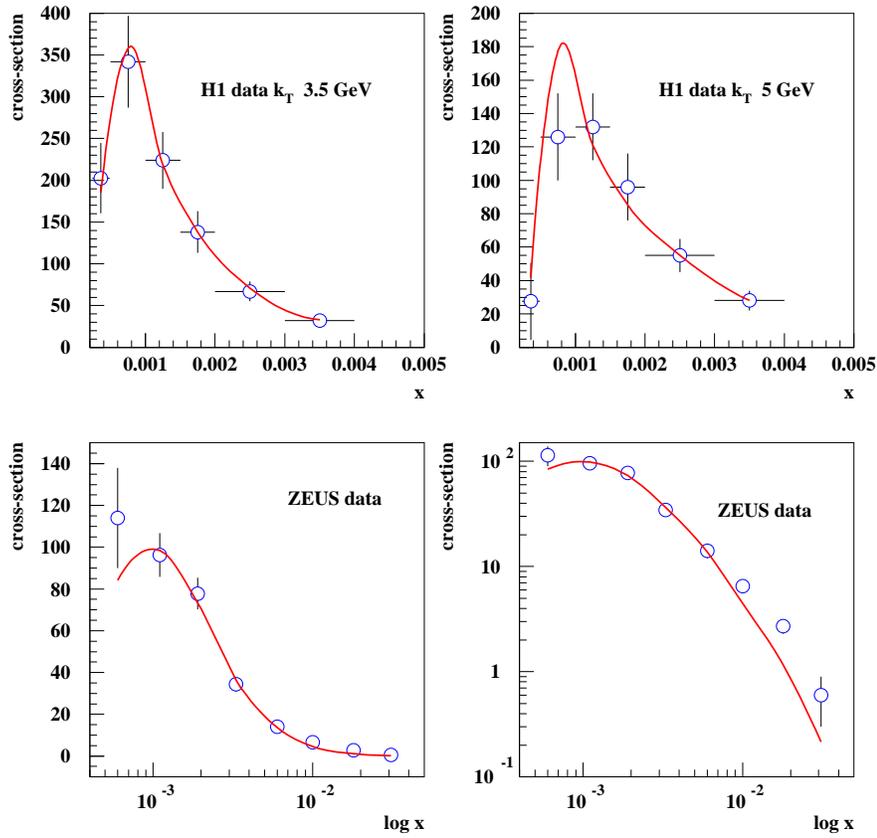,height=5.in}}
\end{center}
\caption{The H1 data ($k_T > 3.5$ GeV, $k_T > 5$ GeV), and the ZEUS data are
compared with the result of the fit. ZEUS data are also displayed in 
logarithmic scales in vertical coordinates to show the discrepancy at high
$x$ values.}
\end{figure}

\section{Comparison with Tevatron results and prospects for LHC}
The final result of our new determination of the effective pomeron
intercept is $\alpha_P=1.43 \pm 0.025$ (stat.) $\pm 0.025$ (syst.).
Our method allows a direct comparison of the intercept 
values with those obtained in other experimental processes, i.e.
$\gamma^* \gamma^*$ cross-sections at LEP \cite{ro99}, jet-jet cross-sections
at Tevatron at large rapidity intervals \cite{D0}, $F_2$ and $F_2^D$ proton
structure function measurements \cite{ab00,na96,mu98}.
Let us first consider the
known determinations of the effective intercepts in $F_2$ and $F_2^D$
measurements at HERA \cite{h1ZEUS}.
It is known that the effective intercept determined in these
measurements is
rather low\footnote{It is interesting to note that the ``hard'' Pomeron
intercept obtained within the framework of two-Pomeron models\cite{la99}
fits with our determination. However our parametrization (\ref{dsigma})
corresponds to only one Pomeron.}(1.2-1.3). This is the reason why these
data
can be both described
by a DGLAP or a BFKL-LO fit \footnote{ Note that in the BFKL
descriptions of these data \cite{na96,mu98}, the effective intercept is
taken to be constant, while
the $Q^2$ dependence comes from the BFKL integration (see  for instance
formula (\ref{dsigma}))}.

Now let us consider processes initiated by two hard probes which allow
a more direct comparison between experiments and BFKL predictions.
These processes suppress DGLAP evolution by selecting events
with comparable hard scales for both hard probes.
Recent data on $\gamma^* \gamma^*$ cross-section measurements at LEP
\cite{opall3} lead to a BFKL description with a low effective intercept
compatible with the one of $F_2$ and $F_2^D$ at HERA
($\alpha_P$=1.2-1.3 \cite{ro99}) \footnote{The statistics for these data
is still very low. L3 and OPAL Collaborations have released the cuts
used
to enhance BFKL effects to get more statistics \cite{opall3,ro99}.
These data can be
both described by BFKL and DGLAP evolution equations.}.
The fact that similar values of the intercepts are found could be
interpreted
by sizeable higher order corrections to BFKL equation.
On the other hand,
it is interesting to note that our
result based on forward jet measurement at HERA obtained in comparable
$Q^2$ ($Q^2 \sim 10$ GeV$^2$) and rapidity
($Y \sim$ 3-4) domains is quite different. The value of the intercept is
significantly higher.

Let us compare our results with the effective intercept 
we obtain from recent preliminary dijet data obtained by the
D0 Collaboration at Tevatron \cite{D0}. The measurement consists
in the ratio $R=\sigma_{1800}/ \sigma_{630}$ where $\sigma$ is the dijet 
cross-section at large rapidity interval $Y \sim \Delta \eta$ 
for two center-of-mass energies
(630 and 1800 GeV), $\Delta \eta_{1800}=4.6$, $\Delta \eta_{630}=2.4.$
The experimental measurement is $R=2.9 \pm 0.3$ (stat.) $\pm 0.3$ (syst.).
Using the Mueller-Navelet formula \cite{mu86}, this measurement allows us to get
a value 
of the effective intercept for this
process 
\begin{eqnarray}
R=\frac{\int_{\frac 12- i\infty}^{\frac 12+ i\infty} \frac{d \gamma}{2i \pi
\gamma (1-\gamma)}
e^{\epsilon (\gamma,0) \Delta \eta_{1800}}} 
{\int_{\frac 12- i\infty}^{\frac 12+ i\infty} \frac{d \gamma}
{2i \pi \gamma (1-\gamma)}
e^{\epsilon (\gamma,0) \Delta \eta_{630}}}. 
\label{muelnav}
\end{eqnarray}
We get $\alpha_P$=1.65 $\pm$ 0.05 (stat.) $\pm$ 0.05 (syst.), in agreement with
the value obtained by D0 using a saddle-point approximation \cite{D0}
(see Table 1).
This intercept is higher than the one obtained in the forward jet study.

Formula (\ref{muelnav}) is obtained after integration over 
the jet tranverse energies at 630 and 1800 GeV, $E_{T_1}$, $E_{T_2}$. 
We note that the non integrated formula
\begin{eqnarray}
R(E_{T_1}/E_{T_2})=
\frac{\int_{\frac 12- i\infty}^{\frac 12+ i\infty} \frac{d \gamma}{2i \pi}
\left( \frac{E_{T_1}}{E_{T_2}}\right)^{2 \gamma}
e^{\epsilon (\gamma,0) \Delta \eta_{1800}}} 
{\int_{\frac 12- i\infty}^{\frac 12+ i\infty} \frac{d \gamma}{2i \pi}
\left( \frac{E_{T_1}}{E_{T_2}}\right)^{2 \gamma}
e^{\epsilon (\gamma,0) \Delta \eta_{630}}}
\label{new}
\end{eqnarray}
shows a sizeable dependence on $E_{T_1}/E_{T_2}$, which could be confronted
with experiment. Let us show both the integrated and $E_{T_1}/E_{T_2}$
dependent cross-sections in Figure 2.

Prospects at LHC are quite appealing for this measurement due to the large
rapidity intervals which can be reached. For instance, we estimate that it
is possible to reach a rapidity interval 
$\Delta \eta_{14000}$ of the order of 10 for a center-of-mass
energy of 14 TeV. Taking as a reference value the maximal value at Tevatron
($\Delta \eta_{2000} \sim 5$), which could be also reached at LHC by
reducing the beam energies, we use Formulae (\ref{muelnav},
\ref{new}) with these
values of rapidity ranges to get a prediction on the $R$ values (see figure 3).
The BFKL prediction gives a high value of $R$ ($R \sim 20$), with a 
typical dependence
on $E_{T_1}/E_{T_2}$ quite smaller than at Tevatron, which would be nice to test
at LHC. If we consider two more close rapidity intervals (e.g. 10 and 8
corresponding to center-of-mass energies of 14 and 7 TeV), we get similar
values of $R$ compared to Tevatron ($R \sim 3.2$). However, the 
$E_{T_1}/E_{T_2}$ dependence is much smaller than in Figure 2 for Tevatron.
It favors the BFKL dynamics due to the high rapidity domains involved.
It is thus important to perform this measurement and BFKL test at LHC.

The question arises
to interpret the different values of the  effective intercept for the
different experimental processes. It could
reasonably
come from the differences in higher 
order QCD corrections for the BFKL kernel and/or in the impact factors
depending on the initial probes ($\gamma^*$
vs. jets).
In order to evaluate the approximate size of the higher order BFKL corrections,
we will use their description in terms of rapidity veto effects \cite{nlo}.
In formula (\ref{dsigma}), we make the following replacement 
\begin{eqnarray}
 \exp (\epsilon (\gamma,0) Y) \rightarrow 
\Sigma _{n=0}^{\infty}~ \theta (Y-(n+1)b)~ \frac{\left[ \epsilon(\gamma,0)
~(Y-(n+1)b)
\right]^n}
{\Gamma(n+1)}~. 
\end{eqnarray}
The Heaviside function $\theta$ ensures that a BFKL ladder of
$n$ gluons occupies $(n+1)b$ rapidity interval where $b$ parametrises the
strength of NLO
BFKL corrections. The value of the leading order intercept is fixed to 
$\alpha_p=1.75 (\alpha_S(Q^2=10)=0.28)$, where $Q^2=10$ GeV$^2$ is
inside the average range of $Q^2$ in the forward jet measurement. 
The fitted value of the $b$ parameter obtained
using the forward jet data is found to be 1.28 $\pm$ 0.08 (stat.) $\pm$ 0.02
(syst.).
Imposing the same value of $\alpha_P$ with Tevatron data gives
$b$=0.21 $\pm$ 0.11 (stat.) $\pm$ 0.11 (syst.). Note that the theoretical value 
of $b$ for the NLO BFKL kernel is expected to be of the order 2.4, which is also
compatible with 
the result obtained
for the $\gamma^* \gamma^*$ cross-section. A contribution from the NLO impact
factors is not yet known, and could perhaps explain the different values of $b$.
The LHC would be also a very interesting testing ground for the study of
rapidity veto and higher order BFKL effects.

\begin{figure}
\begin{center}
\centerline{\psfig{figure=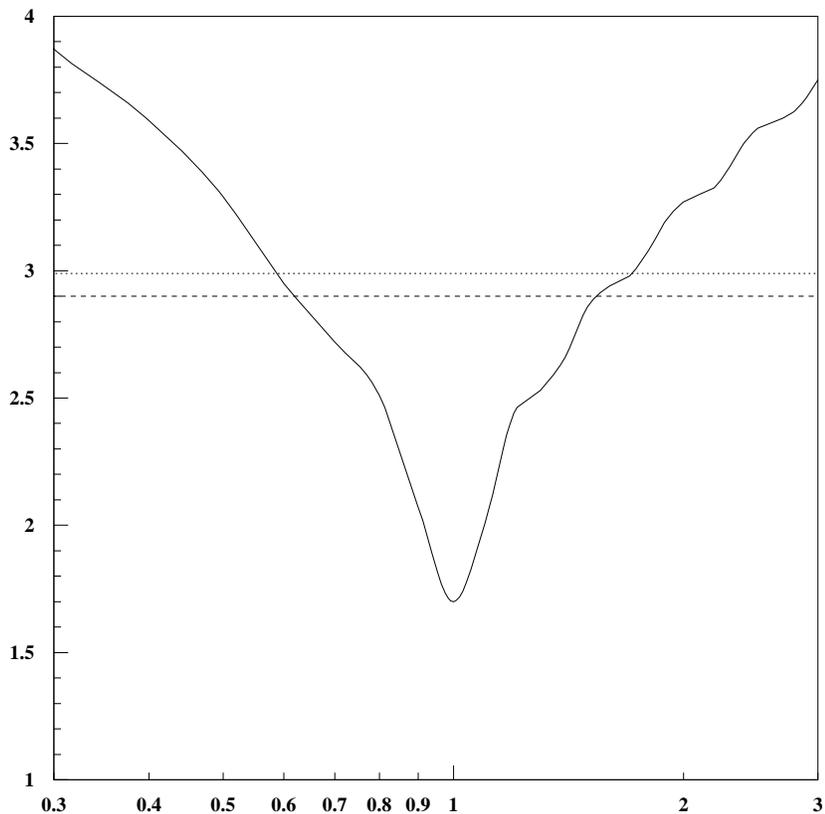,height=5.in}}
\end{center}
\caption{$E_{T_1}/E_{T_2}$ dependence of the dijet cross-section ratio
at Tevatron. $E_{T_1}/E_{T_2}$ is given in the horizontal axis,
and $R$ in vertical axis. In full line is given the non integrated 
$R(E_{T_1}/E_{T_2})$ (see
formula (\ref{new})), in dotted line, the integrated $R$ (formula
(\ref{muelnav}) and in dashed line, the saddle point approximation of $R$
\cite{mu86}, for the fitted value of $\alpha_P$ (see table 1).}
\end{figure}

\begin{figure}
\begin{center}
\centerline{\psfig{figure=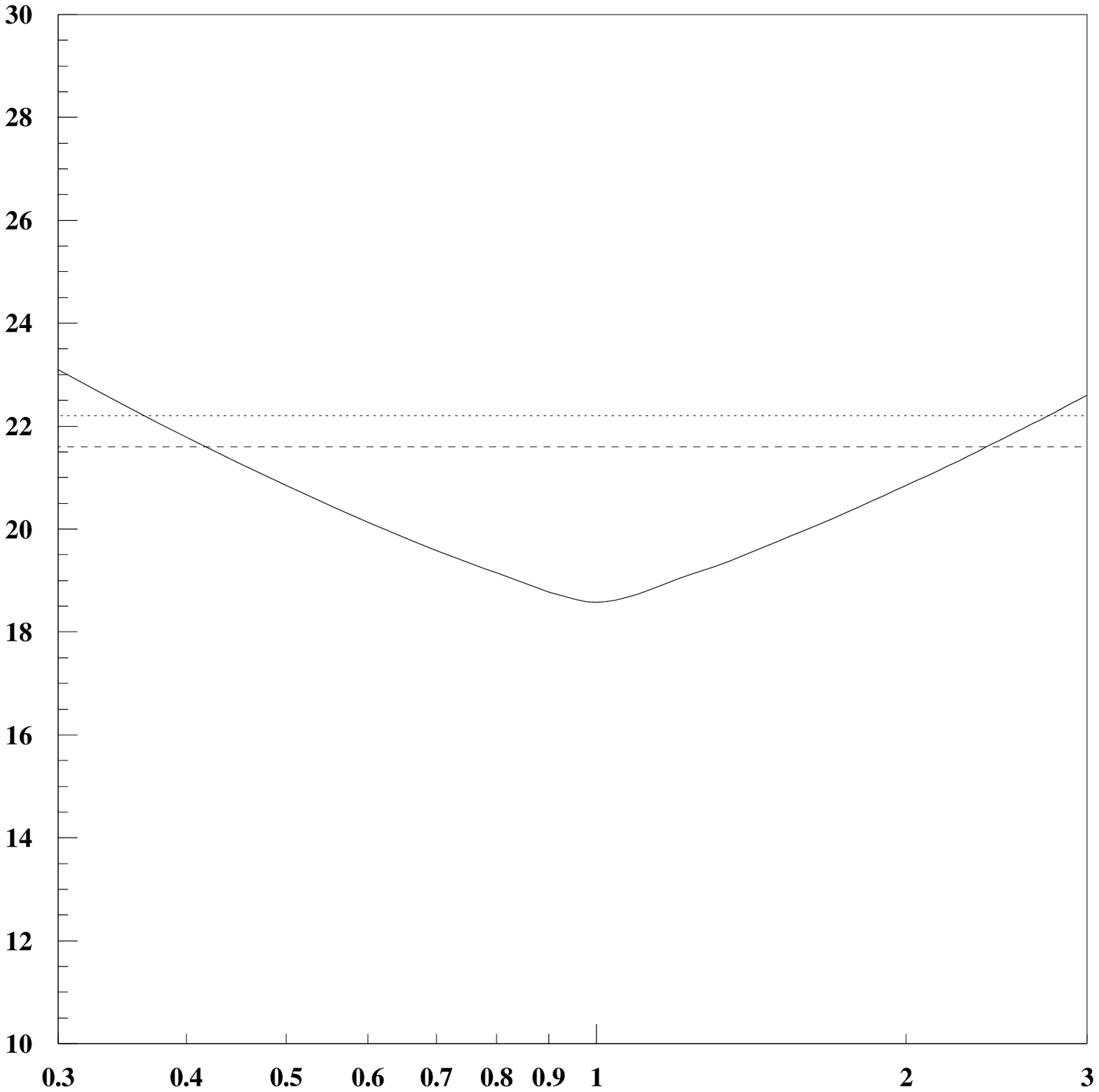,height=5.in}}
\end{center}
\caption{$E_{T_1}/E_{T_2}$ dependence of the dijet cross-section ratio
at LHC for two values of $\Delta \eta$ ($\Delta \eta $(14TeV) $\sim$
10, $\Delta \eta $(2TeV) $\sim$ 4.6).
Same curves and axis definitions as in Figure 2.}
\end{figure}

\section{Conclusion}

To summarize our results, using a new method to disantangle the effects of the
kinematic cuts from the genuine dynamical values 
we find that the effective pomeron intercept of the forward jet
cross-sections at HERA is $\alpha_P=1.43 \pm
0.025$
(stat.) $\pm 0.025$ (syst.). It is much higher than the soft pomeron intercept,
and, among those determined in hard processes,  it is intermediate 
between $\gamma^* \gamma^*$
interactions at LEP and dijet productions with large rapidity intervals at
Tevatron, where we get $\alpha_P$=1.65 $\pm$ 0.05 (stat.) $\pm$ 0.05 (syst.).

Looking for an interpretation of our results in terms of higher order BFKL 
corrections expressed by rapidity gap vetoes $b$ between emitted gluons, we find
a value of $b=$1.3 at HERA, and 0.21 at Tevatron.
The HERA value is sizeable but less than the theoretically predicted
\cite{li98}
value for the NLO BFKL kernel ($b=$2.4).  The Tevatron value is compatible 
with zero.
The observed dependence in the process
deserves further more precise studies \cite{lipatov}.

The LHC will open new possibilities to test different aspects
of BFKL dynamics including higher order effects. Due to the large
ranges in rapidity, large ratios of dijet cross-sections can be expected
by using two different center-of-mass energies. We suggest to measure 
the dependence of the dijet cross-sections as a function of the jet
transverse energies as a signal for BFKL pomeron at Tevatron run II, and at
LHC. 
The Mueller Navelet jet study would also benefit from a lower energy run
at LHC to allow a normalisation independence of the
intercept determination and BFKL tests. LHC will thus allow a measurement
of the pomeron intercept in a different kinematical domain more suited
for BFKL dynamics, and a direct test of higher order effects.


\begin{thebibliography}{9}
\bibitem{mu91}  A.H.Mueller, {\it Nucl. Phys.} {\bf B} (Proc. Suppl.) 18C (1991)
125.

\bibitem{mu86} A.H.Mueller and H.Navelet,  {\it Nucl. Phys.} {\bf B282} (1987) 
107.

\bibitem{D0} A.Goussiou, for the D0 collaboration,  {\it Dijet Cross section at
large 
$s/Q^2$ in $\bar p p$ Collisions}, presented at the `International Europhysics 
Conference On High-Energy Physics' (EPS-HEP 99), Tampere, Finland, July, 1999.

\bibitem{ro99}
S.Brodsky, V.S.Fadin, V.T.Kim, L.N.Lipatov, G.B.Pivovarov,
{\it JETP Lett.} {\bf 70} (1999) 155 , M.Boonekamp. 
A.De Roeck, C.Royon, S.Wallon, {\it Nucl.Phys.} {\bf B555} (1999) 540,
for a recent review and references, Ch. Royon, {\it BFKL signatures at a linear 
collider}, invited talk given at the  International Workshop on Linear Colliders
(LCWS99), April 28- May 5, Sitges (Spain),
hep-ph/9909295. 



\bibitem{al77}  G.Altarelli and G.Parisi,
{\it Nucl. Phys.} {\bf B126}  18C (1977) 298.
V.N.Gribov and L.N.Lipatov, {\it Sov. Journ. Nucl. Phys.} (1972) 438 and 675.
Yu.L.Dokshitzer, {\it Sov. Phys. JETP.} {\bf 46} (1977) 641.

\bibitem{li77}  L.N.Lipatov, {\it Sov. J. Nucl. Phys.} {\bf 23} (1976) 642;
V.S.Fadin, E.A.Kuraev and L.N.Lipatov, {\it Phys. lett.} {\bf B60} (1975)
50; E.A.Kuraev, L.N.Lipatov and V.S.Fadin, {\it Sov.Phys.JETP} {\bf 44}
(1976) 45, {\bf 45} (1977) 199; I.I.Balitsky and L.N.Lipatov, {\it %
Sov.J.Nucl.Phys.} {\bf 28} (1978) 822.

\bibitem{h199} H1 Collaboration, C.Adloff et al. {\it Nucl. Phys.} {\bf B538}
(1999) 
3.

\bibitem{ze99} ZEUS Collaboration, J.Breitweg et al. {\it Eur. Phys. J.} {\bf
C6} 
(1999) 239.

\bibitem{ourpap} G.Contreras, R.Peschanki, C.Royon, hep-ph/0002057,
to appear.

\bibitem{ba92} J.Bartels, A.De Roeck, M.Loewe,
{\it Zeit. f\"ur Phys.} {\bf C54} (1992) 921; 
W-K. Tang, {\it Phys. lett.} {\bf B278} (1992) 635; 
J.Kwiecinski, A.D.Martin, P.J.Sutton, {\it Phys.Rev.} {\bf D46} (1992) 921.

 


\bibitem{li98}
V.S. Fadin and L.N. Lipatov {\it Phys. Lett.} {\bf B429} (1998)127.
M. Ciafaloni {\it Phys. Lett.} {\bf B429} (1998) 363.
M. Ciafaloni and G. Camici {\it Phys. Lett.} {\bf B430} (1998) 349.


\bibitem{ab00}
H.Abramowicz, {\it Diffraction and the Pomeron}
Contribution to the 19th International Symposium on Lepton and Photon
Interactions at 
High-Energies (LP 99), Stanford, California, 9-14 Aug 1999, 
 hep-ph/0001054.
  

\bibitem{na96}
H. Navelet, R. Peschanski, Ch. Royon, {\it Phys. Lett.} {\bf B366} (1995) 329.
H. Navelet, R. Peschanski, Ch. Royon, S. Wallon, {\it Phys. Lett.} 
{\bf B385} (1996) 357.


\bibitem{mu98}
S.Munier, R.Peschanski, Ch.Royon, {\it Nucl.Phys.}
{\bf B534} (1998) 297.



\bibitem{ca91}
S. Catani, M. Ciafaloni, F. Hautmann, {\it Nucl. Phys}. {\bf B366} (1991) 135.
J. C. Collins, R. K. Ellis, {\it Nucl. Phys.} {\bf B360} (1991) 3. E. M. Levin, 
M. G. Ryskin, Yu. M. Shabelskii, A. G. Shuvaev,
{\it Sov. J. Nucl. Phys.} {\bf 53} (1991) 657.

\bibitem{bj71}
J.D.Bjorken, J.Kogut and Soper, {\it Phys.Rev.} {\bf D3} (1971) 1382.
N.N.Nikolaev, B.G.Zakharov, {\it Zeit. f\"ur. Phys.} {\bf C49} (1991) 607;
{\it Phys. Lett.} {\bf B332} (1994) 184.

\bibitem{mu94}
A.H.Mueller, {\it Nucl. Phys.} {\bf B415} (1994) 373;
A.H.Mueller and B.Patel, {\it Nucl. Phys.} {\bf B425} (1994)
471;
A.H.Mueller, {\it Nucl. Phys.} {\bf B437} (1995) 107.


\bibitem{h1ZEUS}
H1 coll., {\it Nucl.Phys.} {\bf B470} (1996) 3, ZEUS Coll., {\it
Z.Phys.} {\bf
C72} (1996) 399, H1 coll., {\it Z.Phys.}
{\bf C76} (1997) 613,
ZEUS
coll., {\it Eur. Phys. J.} {\bf C1} (1998) 81.

\bibitem{la99}
A. Donnachie, P.V. Landshoff {\it Phys.Lett.} {\bf B437} (1998) 408,
{\bf B470}

(1999) 243, and hep-ph/9912312.

\bibitem{opall3}
L3 Coll., {\it Phys. Lett.} {\bf B 453} (1999) 94,
M.Przybycien, OPAL Coll., contribution to the PHOTON99 conference,
Freiburg, Germany, 23-27 May 1999.




\bibitem{nlo} 
L.N.Lipatov, talk presented at the 4th Workshop on Small-x and 
Diffractive Physics, FNAL, September 1998,
C.Schmidt, {\it Phys.Rev.} {\bf D60} (1999) 074003, J.Forshaw, D.A.Ross,
A.Sabio Vera, {\it Phys.Lett.} {\bf B455} (1999)
273-282.

\bibitem{lipatov} 
V.T.Kim, L.N.Lipatov, R.Peschanski, G.Pivovarov, C.Royon, in progress. 


\end{thebibliography}
\end{document}